\documentclass[twocolumn,showpacs,preprintnumbers,amsmath,amssymb  amssymb,
   aps,
   prl,
   lengthcheck,
]{revtex4-1}

\usepackage[english]{babel}

\usepackage{graphicx}
\usepackage{dcolumn}
\usepackage{bm}
\usepackage{hyperref}
\usepackage{makeidx}
\makeindex
\def\>{\right\rangle}
\def\<{\left\langle}
\def\be{\begin{equation}}
\def\ee{\end{equation}}
\def\ba{\begin{array}{l}}
\def\ea{\end{array}}

\def\beq{\begin{eqnarray}}
\def\eeq{\end{eqnarray}}

\usepackage[usenames,dvipsnames]{color}

\begin{document}

\title{Numerical determination of OPE coefficients in the 3D Ising model from
off-critical correlators.}
\author{M. Caselle$^{1}$, G. Costagliola $^{1}$, N. Magnoli$^{2}$}
   \affiliation{
$^1$ Dipartimento di Fisica, Universit\`a di Torino and INFN, Via P. Giuria 1,
10125, Torino,
Italy.\\
$^2$ Dipartimento di Fisica, Universit\`a di Genova and INFN, Via Dodecaneso 33, 16146,
Genova, Italy.
}
\date{\today}

\begin{abstract}
We propose a general method for the numerical evaluation of OPE coefficients in
three dimensional Conformal Field Theories based on the study of the conformal
perturbation of two point functions in the vicinity of the critical point. 
We test our proposal in the three dimensional Ising Model, looking at the magnetic
perturbation of the $<\sigma (\mathbf {r})\sigma(0)>$, $<\sigma (\mathbf
{r})\epsilon(0)>$ and  $<\epsilon (\mathbf {r})\epsilon(0)>$ correlators 
from which we extract the values of $C^{\sigma}_{\sigma\epsilon}=1.07(3)$ and
$C^{\epsilon}_{\epsilon\epsilon}=1.45(30)$. Our estimate for
$C^{\sigma}_{\sigma\epsilon}$ agrees with those recently obtained using conformal
bootstrap methods, while $C^{\epsilon}_{\epsilon\epsilon}$, as far as we know, is
new and could be used to further constrain 
conformal bootstrap analyses of the 3d Ising universality class.

\end{abstract}

\maketitle

\section{Introduction}

In the past few years a renewed interest in the Conformal Bootstrap approach
\cite{Polyakov:1974gs} to three dimensional Conformal Field Theories (CFT) led
   to a set of remarkable results~\cite{Rattazzi:2008pe}-\cite{El-Showk:2014dwa} on
the universality classes of several 3d statistical models.
Among these models a particular attention has been devoted to the 3d Ising case,
both for its physical relevance and for the fact
that in this case very precise numerical estimates exist for the
      scaling dimensions \cite{Martin2010}, which allowed highly non trivial tests of
the conformal
bootstrap results. It would be interesting to perform a similar comparison also for
the Operator Product Expansion (OPE) 
coefficients \cite{Wilson:1969zs}, but for these constants  Monte Carlo estimates with an
accuracy comparable with that of the scaling dimensions are still lacking.
As an attempt to fill this gap we propose in this paper 
a general strategy for the numerical estimate of the structure constants and, as a
proof of concept, we 
evaluate $C^{\sigma}_{\sigma\epsilon}$ and
$C^{\epsilon}_{\epsilon\epsilon}$ in the Ising case. Our estimates should be
considered mainly as an exploratory study on the effectiveness 
of the procedure with a limited use of computer power. We plan in a forthcoming paper to obtain
more precise estimates for these and other OPE coefficients.

It is well known that estimating structure constants is 
much more difficult than estimating scaling dimensions. The most
   naive approach  would be to extract them directly from three point functions (i.e. the
connected correlators of three operators) at criticality. However it is easy to see that
    the subtraction in the connected correlator of the two point components leads to
a mix of contributions with similar exponents which makes it very difficult to
extract the 
sought for OPE coefficients.
In this letter we propose an alternative method based on conformal perturbation
which allows to overcome this problem.
The main idea
is to study the perturbed two point functions using the OPE 
to rewrite them as sums of perturbed one point functions
and then use the known scaling behaviour of one point functions and structure
constants to write the perturbed two point function as an expansion in powers
   of the scaling variable. 
   Following
\cite{Wilson:1969zs} it is always possible to write the short distance behaviour of 
two-point correlators in the vicinity of a
critical point  as a short distance expansion:

\begin{equation}
<O_i ({\bf r} ) O_j ({\bf 0} )>_h = \sum _{k} C_{i j}^k (r;h) <O_k ({\bf 0})>_h
\end{equation}

\noindent
where $<..>_h$ denotes the expectation values with respect to the perturbed action,
the $O_i$ represent a complete set of operators of the conformal theory and the
$C_{i j}^k(r,h)$ are the Wilson coefficients.
While the one point functions $<O_k ({\bf 0})>_h$ cannot be determined from the  knowledge of the
critical correlators and take care of the non-perturbative physics of the model,  the Wilson coefficients
can be obtained from integrals of critical correlators.
Moreover it was shown in  \cite{Guida:1995kc}  that the $n^{th}$ order derivatives of the
Wilson coefficients with respect
to the perturbing parameter of the theory  are always infrared finite.
The structure constants can then be easily
extracted from the coefficients of the short distance expansion (see \cite{Guida:1995kc,Guida:1996ux,Caselle:1999mg} for further details). In practice one is
limited
    for numerical reasons to the first few coefficients (typically the first two or
three) which however (combining all the possible two points functions and
     relevant perturbations) are enough to obtain 
all the structure constants among relevant operators.

As it is easy to see the method is based on very general features of CFTs and OPEs.
It can be
applied to any d-dimensional CFT 
and in fact it was applied  a few years ago to the 2d Ising model perturbed by a magnetic field
\cite{Guida:1996ux,Caselle:1999mg}, to various perturbations of the Tricritical Ising model \cite{Guida:1996nm} 
and then used to identify the signatures of subleading irrelevant operators in the perturbed two
point functions \cite{Caselle:2001zd}. 
In all these  cases the structure
constants were already known from the exact solution of Belavin
Polyakov and Zamolodchikov~\cite{Polyakov:1984yq} and the main emphasis was on comparing 
the perturbative results with the numerical simulations.

The aim of this paper is to apply this strategy to the three dimensional case. 
In particular we shall address, as an example, the magnetic perturbation of the 3d
Ising model.

\section{Ising model perturbed by a magnetic field $h$}

The action of the continuum Ising model in $ 3D $ perturbed by a magnetic field is:

\begin{equation}
S = S_{cft} + h\int \sigma (\mathbf {r} ) d \mathbf{r}.
\end{equation}

\noindent
where $S_{cft}$ is the action of the conformal field theory which describes the
model at the critical point.

At the conformal point there are two quasi-primary relevant fields $\sigma$ and
$\epsilon$, whose dimensions are given  by
$\Delta_\sigma = 0.51814(5)$, $\Delta_\epsilon = 1.41275(25)$ \cite{Martin2010}.

The VEVs acquire a dependence on $h$, that can be fixed using renormalization group
arguments:
\beq
%\begin{equation}
\label{eq:mv1}
<\sigma> &=& A_{\sigma} h^{\frac{\Delta_\sigma}{\Delta_h}}\\
%\end{equation}
%\noindent
%and
%\begin{equation}
\label{eq:mv2}
<\epsilon> &=& A_{\epsilon} h^{\frac{\Delta_\epsilon}{\Delta_h}}
%\end{equation}
\eeq
\noindent
In order to simplify notations let us introduce
the scaling variable $ t=|h| r^{\Delta_h}$, and an adimensional version of the structure constants and their derivatives:
\beq
C_{ij}^k &\equiv& \lim_{r->\infty} C_{ij}^k(r;0) \, r^{dim(C_{ij}^k(r;0))}\nonumber \\
%\eeq
%\beq
\partial_hC_{ij}^k &\equiv& \lim_{r->\infty} \partial_hC_{ij}^k(r;0)  \, r^{dim(\partial_hC_{ij}^k(r;0))}\nonumber
\eeq
With these definitions the short distance expansion of $\sigma$
and $\epsilon$ correlators becomes:

\beq\label{eq:cor1}
r^{2 \Delta_{\sigma}}<\sigma (\mathbf {r})\sigma(0)> &=& C^1 _{\sigma \sigma} +
A_\epsilon C^{\epsilon}_{\sigma\sigma} t^{\frac{\Delta _\epsilon}{\Delta _h} }\\
&+& A_\sigma \partial_h C^\sigma_{\sigma \sigma} t^{\frac{\Delta_\sigma}{\Delta _h}+1} + O(t^{2}) \nonumber
\eeq

\beq \label{eq:cor3}
r^{\Delta_{\sigma}+\Delta_{\epsilon}}<\sigma (\mathbf {r})\epsilon(0)> &=& A_\sigma
C^{\sigma}_{\sigma\epsilon} t^{\frac{\Delta _\sigma}{\Delta _h} }+  \partial_h
C^1_{\sigma \epsilon} t \\
&+&A_\epsilon
   \partial_h C^\epsilon_{\sigma \epsilon}
t^{\frac{\Delta_\epsilon}{\Delta_h}+1} + O(t^{\frac{\Delta_\sigma}{\Delta _h}+2}) \nonumber
\eeq

\beq \label{eq:cor2}
r^{2 \Delta_\epsilon}<\epsilon (\mathbf {r})\epsilon(0)> &=& C^1 _{\epsilon \epsilon}
+ A_\epsilon C^{\epsilon}_{\epsilon\epsilon} t^{\frac{\Delta _\epsilon}{\Delta _h}} \\
&+& A_\sigma \partial_h C^\sigma_{\epsilon\epsilon} t^{\frac{\Delta_\sigma}{\Delta _h}+1}+ O(t^{2})\nonumber
\eeq

\noindent
where  $\Delta_h = 3 -\Delta_{\sigma} =  2.48186(5) $ is the dimension of the magnetic
field and with the standard CFT normalizations we may set $C^1 _{\epsilon \epsilon}=C^1 _{\sigma \sigma}=1$.

\subsection{Conversion from lattice to continuum normalizations.}

The partition function of the 3d Ising model is given by:

\begin{equation}
Z = \sum_{\sigma _i = \pm 1 } e^ {\beta (\sum _{<i,j>}\sigma _i \sigma _j +H \sum _i
\sigma _i) }
\end{equation}

\noindent
where $<i,j>$ indicates nearest neighbors sites in the lattice  which we assume to
be a three-dimensional cubic lattice of size $L$.
Fixing $\beta$ at its critical value  $\beta _c= 0.2216544 $ \cite{Blote:1996}  and defining $h_l =
\beta _c H$, we have:

\begin{equation}
Z = \sum_{\sigma _i = \pm 1 } e^{ \beta _c\sum _{<i,j>}\sigma _i \sigma _j + h_l
\sum _i \sigma _i }
\end{equation}

\noindent

It is natural to define the lattice discretization of $\sigma$ as $\sigma^{l} \equiv 
\frac{1}{L^3}\sum _i \sigma _i$, so that its mean value
coincides with the magnetization.
In a similar way one can define the energy operator $\epsilon^{l} $ as $\epsilon^{l}
\equiv \frac{1}{3 L^3} \sum_{<i,j>} \sigma _i \sigma _j -\epsilon_{cr}$, where
$\epsilon_{cr}$ is the energy
at the critical point, so that the mean value of $\epsilon^{l}$  coincides with the
singular part of the
internal energy. 
\noindent

It is important to notice that measuring the mean values (\ref{eq:mv1},\ref{eq:mv2}) and the 
two point functions (\ref{eq:cor1},\ref{eq:cor3},\ref{eq:cor2}) on the lattice we 
find the lattice versions of the amplitudes $A$ and structure constants $C$ (which we shall denote in the following with the index $l$: $A^l$ and $C^l$).
To relate these constants with the continuum ones we must first fix the
relative normalization of $\sigma^{l} , \epsilon^{l}$ and $h_l$ with respect to
$\sigma , \epsilon $ and $h$.
The simplest way to do this is by comparing the correlators at the critical point. 
\noindent
In fact from 
\begin{equation} \label{eq:corlat}
<\sigma _i \sigma _j > = \frac{R_\sigma ^2}{| r_{ij} |^{2 \Delta _\sigma}},
\end{equation}
\noindent
we get (assuming the standard normalization of the two point function in the continuum)  
$\sigma^{l} = R_\sigma \sigma$. 
By considering the  $\epsilon^{l}$ correlator we  get similarly $\epsilon^{l}  =
R_\epsilon \epsilon$. Finally, $h_l$ is fixed by matching the lattice and continuum perturbation terms:
$h_l = R_\sigma ^{-1} h$.

Using $R_\sigma$ and $R_\epsilon$ we can easily obtain continuum amplitudes and OPE constants from the lattice ones.
 For example from the equation
~(\ref{eq:mv2}) we get 
$  A_{\epsilon}^{l} = {R_{\sigma}}^{\frac{\Delta \epsilon}{\Delta h}} \;
R_{\epsilon} A_{\epsilon} $ 
and from this and the equations ~(\ref{eq:cor1}) and ~(\ref{eq:corlat}) we have the
rescaling 
for the structure constant: ${C_{\sigma \sigma}^{\epsilon}}^{l}  R_{\epsilon} =
C_{\sigma\sigma }^{\epsilon} $.

\section{Monte Carlo Simulations}

\subsection{Estimating $R_{\sigma}$ and $R_{\epsilon}$}

As a first step we extracted the values of $R_{\sigma}$ and $R_{\epsilon}$ from a finite size scaling analysis of the
two point correlators at the critical point.
   The table below summarizes our results. We are not aware of independent estimates of these quantities except for  $\epsilon_{cr} $ 
   which was estimated in \cite{Martin2012} as $\epsilon_{cr} = 0.3302022(5)$ in good agreement with our result. 
   The quoted uncertainties combine both
statistical and systematic errors. It should be noticed,
   in view of possible future improvements of our analysis, that the main source of
error was due to the infinite volume extrapolation of our results (we used lattices in the range $120\leq L\leq 200$).

   \begin{table}[h] \label{tab:pre}
   \caption{Rescaling and other useful constants }
   \centering

   \begin{tabular}{p{50pt} l}
   \hline
   $R_{\sigma}$                 & 0.550 (4) \\
   $R_{\epsilon}$        & 0.237 (3) \\
   $A_{\sigma}^{l} $        & 1.0125 (5) \\
   $A_{\epsilon}^{l}$        & 0.608 (1) \\
   $\epsilon_{cr} $        & 0.330213 (12)\\
   \hline
   \end{tabular}

   \end{table}

\subsection{Perturbed correlators}

   To extract the structure constants we then estimated the same correlators in
presence of a small magnetic field $h_{l}$.
   The choice of $h_{l}$ is constrained by two main requirements:
   $h_{l}$ should be small enough to keep the correlation length
   $\xi$ as large as possible, at the same time it should not be too small to avoid
finite size effects.
   The optimal choice is thus fixed by the maximal lattice size which we
could simulate with the computer resources at our disposal, which was fixed to be
   $L=200$. Given this constraint the optimal range turned out to be
   $2 \cdot 10^{-5} < h_{l} < 0.75 \cdot 10^{-5}$. In this range the correlation
length spans from about $20$ to $40$ lattice spacings, allowing to sample a
   sufficient number of different distances of the correlators.

   For all values of $h$ we performed our simulations on lattices of size $L=200$ 
   using a state of art Monte Carlo algorithm with
   about $10^{7}$ configurations for each simulation. 
   We fitted the   correlators with  eq.s (\ref{eq:cor1}, \ref{eq:cor3}, \ref{eq:cor2}),   
setting $C^1 _{\sigma \sigma}=C^1 _{\epsilon \epsilon}=1$,
   using $A_{\sigma},A_{\epsilon},\Delta_\sigma,\Delta_\epsilon$  as fixed inputs
and keeping as only free parameters the structure constants. It turned out that
in all
   correlators the last terms quoted in eq.s (\ref{eq:cor1}, \ref{eq:cor2}) and the last two terms in  eq. \ref{eq:cor3} were
negligible within the errors and in all three cases we ended up with a linear fit
with only one
   free parameter. 
   A major source of uncertainty in our estimates is the systematic error due to the uncertainty in the estimates of $R_\sigma$ and 
   $R_\epsilon$. We quote separately in the following these errors (which we report in square brackets) from the
   statistical ones.
Due to the different scaling behaviour, the relative size of statistical and systematic errors is different for the three correlators.
 For $C_{\sigma \sigma }^{\epsilon} $ estimated from $<\sigma (\mathbf {r})\sigma(0)>$  the two errors are of similar magnitude. 
 $C_{\epsilon \epsilon }^{\epsilon} $ extracted from $<\epsilon
(\mathbf {r})\epsilon(0)>$ is dominated by the systematic
error while for $C_{\sigma \epsilon }^{\sigma}$.
 extracted from $<\sigma (\mathbf {r})\epsilon(0)>$ the systematic error is negligible (due to the fact that the critical correlator vanishes in this case).

   The tables below report our results for the structure constants. $r_{min}$ and $r_{max}$ denote the range of distances included in the fit. 
   In all cases we chose $r_{min}=5$. We verified that in all three cases this was enough 
   to eliminate lattice artifacts and to neglect short distance subleading contributions in the correlators.

   \begin{table}[h]
   \caption{Results for the structure constant $C_{\sigma \sigma }^{\epsilon} $
obtained from the spin-spin correlator. We report systematic errors in square brackets. \label{tab:const1}}
   \centering

   \begin{tabular}{p{70pt} p{30pt} p{30pt} p{70pt}  }
   \hline
   $h_{l}$ & $r_{min}$ & $r_{max}$ & $C_{\sigma \sigma }^{\epsilon} $  \\
   \hline
   $ 2 \cdot10^{-5}$           & 6 & 20 & 1.06 (3)[7]         \\
   $ 10^{-5}$                  & 5 & 35 & 1.04 (2)[7]         \\
   $ 0.85 \cdot 10^{-5}$       & 5 & 35 & 1.06 (3)[8]         \\
   $ 0.75 \cdot 10^{-5}$       & 5 & 40 & 1.07 (3)[8]        \\
   \hline
   \end{tabular}

   \end{table}

   \begin{table}[h]
   \caption{Results for the structure constant $C_{\epsilon \epsilon }^{\epsilon} $
obtained from the energy-energy correlator. We report systematic errors in square brackets. }
   \centering

   \begin{tabular}{p{70pt} p{30pt} p{30pt} p{70pt}  }
   \hline
   $h_{l}$ & $r_{min}$ & $r_{max}$ & $C_{\epsilon \epsilon }^{\epsilon}$  \\
   \hline
   $ 2 \cdot 10^{-5}$         & 5 & 16 & 1.47 (10) [30]              \\
   $ 10^{-5}$                 & 5 & 18 & 1.42  (10) [50]               \\
   $ 0.85 \cdot 10^{-5}$ & 5 & 18 & 1.43 (12) [50]             \\
   \hline
   \end{tabular}
   \label{tab:const2}
   \end{table}

   \begin{table}[h]
   \caption{Results for the structure constant $C_{\sigma \epsilon }^{\sigma}$ 
obtained from the spin-energy correlator. }
   \centering

   \begin{tabular}{p{70pt} p{30pt} p{30pt} p{70pt}  }
   \hline
   $h_{l}$ & $r_{min}$ & $r_{max}$ & $C_{\sigma \epsilon }^{\sigma}$ \\
   \hline   
   $ 2 \cdot 10^{-5}$ 	& 5 & 16 & 1.10 (3)         \\
   $ 10^{-5}$                 	& 5 & 16 & 1.07 (4)          \\
   $ 0.85 \cdot 10^{-5}$ 	& 5 & 14 & 1.07 (5)         \\
   \hline
   \end{tabular}
   \label{tab:const3}
   \end{table}

Looking at tab.s (\ref{tab:const1},\ref{tab:const3}) we see that, within our statistical errors 
$C_{\sigma \sigma }^{\epsilon} = C_{\sigma \epsilon }^{\sigma}  $. 
  Combining the results listed in tab.s (\ref{tab:const1},\ref{tab:const2},\ref{tab:const3}) we quote
as our final estimate for the two structure constants
  $C^{\sigma}_{\sigma\epsilon}=1.07(3)$ and
$C^{\epsilon}_{\epsilon\epsilon}=1.45(30)$. The first value is in good
agreement with a recent conformal bootstrap calculation \cite{El-Showk:2014dwa}:
$(C^{\sigma}_{\sigma\epsilon})^2=1.10636(9)$ \cite{Nota1}.
We are not aware
of any estimate for the second constant $C^{\epsilon}_{\epsilon\epsilon}$. The
very fact that it is different from zero is rather non trivial. In fact in the two
dimensional model $C^{\epsilon}_{\epsilon\epsilon}=0$ as a consequence of dual
symmetry.
  Our calculation shows that indeed this is not a generic property of the Ising
model but is a specific feature of the two dimensional self-dual case.

\section{Concluding remarks}
In this paper we proposed a general strategy for the Monte Carlo estimate of OPE
coefficients in d-dimensional spin models. As a proof of concept of our method we
performed an exploratory study in the 3d Ising case and found a value of 
$C^{\sigma}_{\sigma\epsilon}$ in agreement with the known conformal bootstrap
results  and  a preliminary estimate for 
$C^{\epsilon}_{\epsilon\epsilon}$ which could be used as input for further constrain
existing conformal bootstrap calculations. Similarly, the knowledge of 
these structure constants could be used to constrain 
the linear response dynamics at finite temperature of Conformal Quantum
Critical  systems \cite{WWK2015}. 

We see a few possible improvements of our approach:
\begin{itemize}
\item The systematic uncertainty which is the major source of error in
$C^{\epsilon}_{\epsilon\epsilon}$ 
could be improved,  following the approach proposed 
in \cite{Martin2010}, looking at different realizations of the 3d universality class
with improved scaling behaviour.
\item
It would be useful to combine the magnetic perturbation that we studied in this
paper with other types of perturbation. 
\item
It would be interesting to extend the method also to boundary CFTs \cite{Liendo:2012hy} 
and to CFTs containing conformal defects \cite{Billo:2013jda,Gaiotto:2013nva}
\end{itemize} 
We plan to address these issues in a forthcoming publication.

{\bf Acknowledgements}
We thank F. Gliozzi, M. Hasenbusch, A. Rago, S. Rychkov and W. Witczak-Krempa 
for useful discussions and suggestions.

\end{document}